\documentclass[11pt]{article}
\usepackage{amsmath}
\usepackage[margin=2.0cm]{geometry}
\usepackage{epsfig}
\usepackage{amsfonts}
\usepackage{titlesec}
\usepackage{fancyhdr}
\usepackage{setspace}
\usepackage{authblk}
\usepackage{subcaption}
\usepackage{xcolor}
\usepackage{hyperref}
\usepackage{cite}
\usepackage[nolists,nomarkers]{endfloat}

\usepackage{tabularx}
\usepackage{epstopdf}
\usepackage{epsfig}
\usepackage{epstopdf-base}
\newcommand{\runningtitle}{Spherical particle formation in abrasion} 
\newcommand{\authorname}{HS Dhami \emph{et al}.} 
\newcommand{\txq}[1]{\textsl{#1}}
\newcommand{\erf}{\ensuremath{\text{erf}}}

\titleformat{\section}
{\normalfont\bfseries\scshape}{\thesection}{1em}{}

\titleformat{\subsection}
{\normalfont\bfseries\scshape}{\thesubsection}{1em}{}

\titleformat{\subsubsection}
{\normalfont\slshape}{\thesubsubsection}{1em}{}

\titleformat{\title}{\bfseries}{\thesection}{1em}{}

\title{\vspace{-3em} \large \textbf{Of fiery sparks \& glittering spots: Melting-resolidification and spherical particle formation in abrasion} }
\author[1]{\normalsize Harish Singh Dhami}
\author[1]{\normalsize Priti Ranjan Panda}
\author[1]{\normalsize Koushik Viswanathan \thanks{Email: koushik@iisc.ac.in}}
\affil[1]{\normalsize \textsl{Dept. of Mechanical Engineering, IISc, Bangalore}}

\begin{document}
\maketitle
\thispagestyle{plain}
\begin{abstract}
  The curious occurrence of perfectly spherical particles when a steel substrate is slid against a hard abrasive was first observed and documented by Robert Hooke in the 17$^{th}$ century. Similar particles have subsequently been observed in numerous other abrasion-type processes, ranging from grinding of steels to sliding rock faults. The prevalent hypothesis, originally proposed by Hooke, is that these particles are formed due to high local temperatures between the abrasive and the substrate, resulting in melting, droplet ejection and subsequent resolidification---the melting-resolidification hypothesis. In this work, we revisit this phenomenon using \emph{in situ} analysis of a model steel-abrasive contact geometry, complemented by analytical calculations. It is found that the temperature within the contact zone, for typical contact conditions used, is far from the melting point and that spherical particles do not form in the absence of oxygen. We thereby propose a modification of the melting-resolidification hypothesis, involving an intermediate exothermic oxidation stage, and provide quantitative evidence for each step of the process. Our results have implications for a wide class of abrasive systems that involve the formation and utilization of spherical metallic particles.          
\end{abstract}
\hrulefill
\section{Introduction}
The curious occurrence of substantial quantities of perfectly spherical particles during abrasion of a steel surface was first reported and studied by Robert Hooke in his highly influential treatise \cite{HookeMicrographia}. Hooke systematically analyzed the sparks formed when a steel specimen was struck against a piece of flint and found \txq{\lq \ldots certain very small, black, but glittering spots \ldots \rq } that, when examined with his microscope, were
\begin{quotation}
  \txq{\ldots pretty regularly round \ldots being fastened onto the end of a small sliver of Iron, which seemed to be nothing else but a long thin chip \ldots some of these sparks are the slivers of Iron \ldots [others] are only small slivers of the Iron, made red-hot with the violence of the stroke\ldots}
\end{quotation}

\begin{figure}
  \centering
  \includegraphics[width=0.8\textwidth]{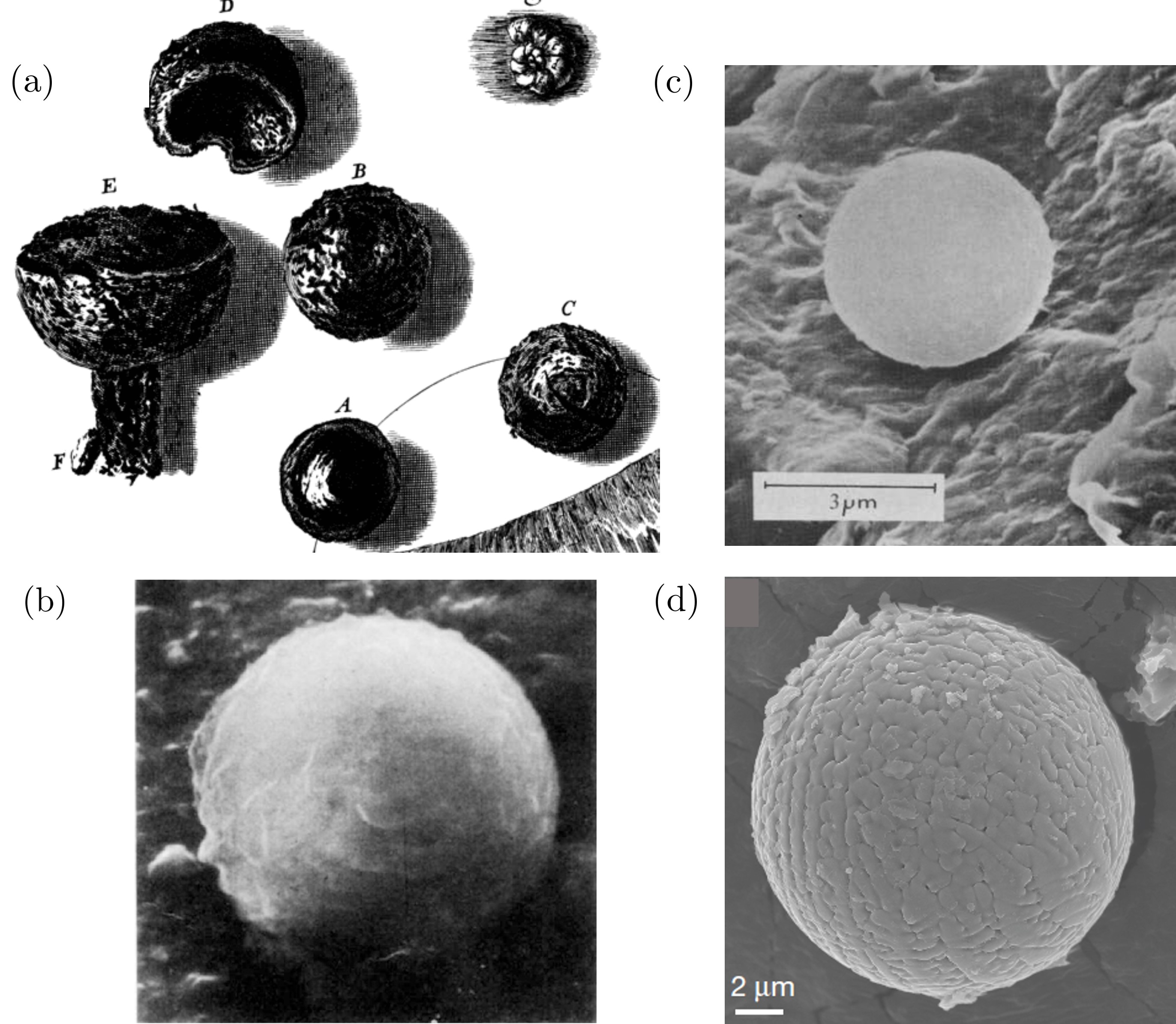}
  \caption{Images showing the occurrence of spherical in different physical contexts (a) Hooke's original drawings showing magnified images of the \lq glittering spots\rq\ observed when rubbing steel and flint \cite{HookeMicrographia}. (b) Spherical wear particle formed due to repeated burnishing \cite{ScottMills1973b}. (c) Spherical particles recovered from sliding rock fault \cite{swain1976wear} and (d) Micrometeorite sample showing perfect spherical shape and dendritic surface patterns \cite{TomkinsETAL2016}.}
  \label{fig:sphere_applications}
\end{figure}

Based on these detailed observations, Hooke put forth a hypothesis on how these spherical particles were formed and, indeed, on the nature of the spark itself. In modern parlance, he essentially proposed that friction between the flint and steel sample, being operative at local asperity contacts, caused large local temperature rise (\lq \txq{red-hot}\rq ). Consequently, he reasoned, the sparks themselves were nothing but these locally hot regions removed in the form of slivers or chips of iron. In certain cases, depending on the size of the chip, the inherently high local temperatures were postulated to cause complete melting, thereby showing up as round \lq \txq{glittering spots}\rq\ post solidification, see Fig.~\ref{fig:sphere_applications}(a). He supported this proposal by drawing attention to common knowledge---that pieces of violently rubbed steel can indeed reach temperatures high enough to cause skin damage.

Subsequent to these ingenious observations, sub-millimeter sized spherical particles have appeared in numerous abrasion problems ranging from conventional wear to sliding fault planes. A series of papers in the seventies established the now-common occurrence of spherical wear debris in rolling contact fatigue \cite{ScottMills1973a, ScottMills1973b, LoyMcCallum1973} and fretting wear of sliding surfaces \cite{Hurricks1974, StowersRabinowicz1972}. These systems certainly could not accommodate material melting since temperatures in lubricated contacts seldom exceed the boiling point of the lubricant ($\sim 300 ^\circ$C) \cite{Rabinowicz1977}. Surface damage in such contacts likely occurs during repeated sliding of the two surfaces, either by adhesive wear or by sub-surface cracking \cite{WaterhouseTaylor1974}, leading to the removal of a small particle from the surface and suspended/transported by the lubricant. This particle is then trapped locally within the contact (perhaps inside a cavity) and is repeatedly burnished by asperities on the two sliding surfaces or even other, larger and potentially harder, wear particles. The result is a near-perfect spherically shaped particle ejected from the surface, see Fig.~\ref{fig:sphere_applications}(b). The process described here is not unlike that seen in the rounding of pebbles by abrasive-laden river water flow \cite{Bloore1977, NovakETAL2018}. We will henceforth refer to this mechanism as abrasion-burnishing and it represents an alternative route to Hooke's melting-resolidification process.

On much larger scales, spherical wear particles are also observed on rock faces in the vicinity of active fault planes \cite{swain1976wear}, a seemingly scaled--up version of the abrasion wear problem described above, see Fig.~\ref{fig:sphere_applications}(c). Here the fault plane constitutes the interface, with two sliding (albeit very slowly) rock bodies on either side. The possibility that these spherical wear particles form via the abrasion-burnishing mechanism---involving trapped rock debris burnished into spherical shape---appears to be readily plausible. However, this system is distinguished by the fact that it represents an unlubricated contact so that significant interface heating is to be expected. In fact, analytical estimates suggest that local temperatures can often exceed the material's melting point \cite{MckenzieBrune1972}. Thus, despite its similarity with the sliding wear system, it is very likely that spherical particles in the fault system form via the Hooke melting-resolidification mechanism rather than abrasion-burnishing. An analogous example is seen in cavitation erosion, where large instantaneous local temperatures can result in melting, particle ejection and resolidification \cite{HaoshengETAL2008, FlintSuslick1991}. It is clear from these instances that in order to ascertain which of the two mechanisms are operative, one requires detailed understanding of the process thermomechanics at multiple scales---from asperity contacts to large scale heat dissipation.

That the melting-resolidification route is a generic and more common mechanism is also borne out by its appearance in yet another geological phenomenon---micrometeorite dust, see Fig.~\ref{fig:sphere_applications}(d). Early observations of dust collected from meteor showers revealed the presence of so-called cosmic spherules \cite{Landsberg1947}, formed due to frictonal heating, breakup and melting of larger micrometeoroid objects during atmospheric entry \cite{GengeETAL2008}. A simple energy balance calculation reveals the maximum size of spherules to be less than $10\,\mu$m \cite{WhipplePNAS1950}. Cosmic spherules are geologically important because their occurrence and chemical composition are linked with their travel history, enabling back-calculation of the composition of the earth's atmosphere during previous geological eras \cite{LoveBrownlee1991}. Here the solidification route is, conceptually, much more obvious, as is the evidence in support of it \cite{TomkinsETAL2016, PayneETAL2020}.

Taken \emph{in toto}, these observations appear to suggest that the melting-resolidification mechanism is far more widespread than the abrasion--burnishing route. However, while this mechanism has been well established in the contexts stated above, its utility in explaining Hooke's original observations is ironically plagued by two fundamental problems. Firstly, more recent studies have shown that local temperatures in dry sliding contact simply cannot exceed the melting point, especially under the pressures and shear stresses normally associated with rubbing a ductile metal (steel) against an abrasive (flint) \cite{shen2008thermocouple, malkin2007thermal, Shaw1996, hwang2003measurement}. This raises an important objection of how melting can occur in the first place. Secondly, the correlation between the occurrence of a spark and the existence of a spherical particle is beyond doubt and suggestive---so much so that in a nitrogen atmosphere, neither sparks nor any spherical particles are observable \cite{LingLuThesis, LingLuETAL1992}. This strongly hints at the important role played by oxygen in effecting the formation of spherical particles, a fact that few proponents of the melting-resolidification mechanism have explicitly accounted for (see, for instance, Refs. \cite{KleisETAL1979, KomanduriShaw1975, RamanathETAL1987}). Finally, indubitable evidence for the resolidification mechanism has hitherto not been provided. This process must happen rapidly and, consequently, leave its own signature in the final spherical particle's properties.

In this light, we revisit the melting-resolidification mechanism for the specific problem of an abrasive (flint) removing material from a ductile substrate (steel). Our experiment uses an abrasive wheel and a sliding steel sample which, apart from being scientifically controllable, also closely resembles Hooke's original configuration, see Sec.~\ref{sec:experimental}. We put forth a modified melting-resolidification hypothesis in Sec.~\ref{sec:hypothesis} and present experimental evidence and accompanying analytical calculations in support of this hypothesis in Sec.~\ref{sec:results}. This latter section follows the life of a single metal chip as it is ejected from the contact zone. Our results and their implications are discussed in Sec.~\ref{sec:discussion} and concluding remarks presented in Sec.~\ref{sec:conclusions}.

\section{Experimental configuration}
\label{sec:experimental}

A surface grinding configuration was used to generate spherical particles, and is analogous to Hooke's original steel-flint abrasion experiments \cite{Shaw2005}. The workpiece and wheel were AISI 52100 high carbon steel and alumina abrasive wheel of 170 mm diameter, respectively. Spindle-wheel angular velocity was kept constant at 2800 rpm. The workpiece was held over an auto-feed bed with controlled translation velocity. Material ejected from the wheel-workpiece contact zone (grinding swarf) was collected at different distances using a magnetic container. Experiments were performed without the magnetic container as well to ensure that there were no magnetic field effects on the particles produced. No coolant was used during the grinding process. The following parameters were used for the experiments, unless specified otherwise: wheel velocity 30 m/s, workpiece (bed-translation) velocity 0.15 m/s, nominal depth of wheel engagement 100 $\mu$m, with a workpiece contact width of 10 mm. Simultaneous \emph{in situ} temperature measurements were performed using an infra-red camera (Optris 640PI) at a framing rate of 32-125 frames per second (FPS). 

In order to isolate spherical particles, the collected powder was sieved using both 45 $\mu$m and 75 $\mu$m mesh sizes. For \emph{post mortem} analyses, the following systems were employed. A FESEM system (Zeiss Ultra55) with electron dispersive spectroscopy (EDS) was used for imaging and elemental analysis. X-ray diffraction of the bulk sample before grinding and of the collected particles was performed using Cu-K$\alpha1$(1.543\AA) source with scan rate 1$^\circ$ per minute. A focused ion beam (FIB) with a micro-manipulator was used to section individual spherical particles and make samples for transmission electron microscopy (TEM). A sample of dimensions $8\times 15\times 5 \,\mu$m was sliced radially from the spherical particle and thinned down to 40 nm thickness using FIB milling. TEM and EDS were performed on the prepared sample. High-resolution transmission electron microscopy (HRTEM) images were analyzed using GMS-3 DigitalMicrograph software.

\section{Modified melting-resolidification hypothesis}
\label{sec:hypothesis}

We first present three figures showing the nature, size and morphology of particles produced during the abrasion process described in Sec.~\ref{sec:experimental}. Based on this, and prior to detailed analysis, we propose a modified melting-resolidification hypothesis to explain their formation.

\begin{figure}
  \centering
  \includegraphics[width=0.8\textwidth]{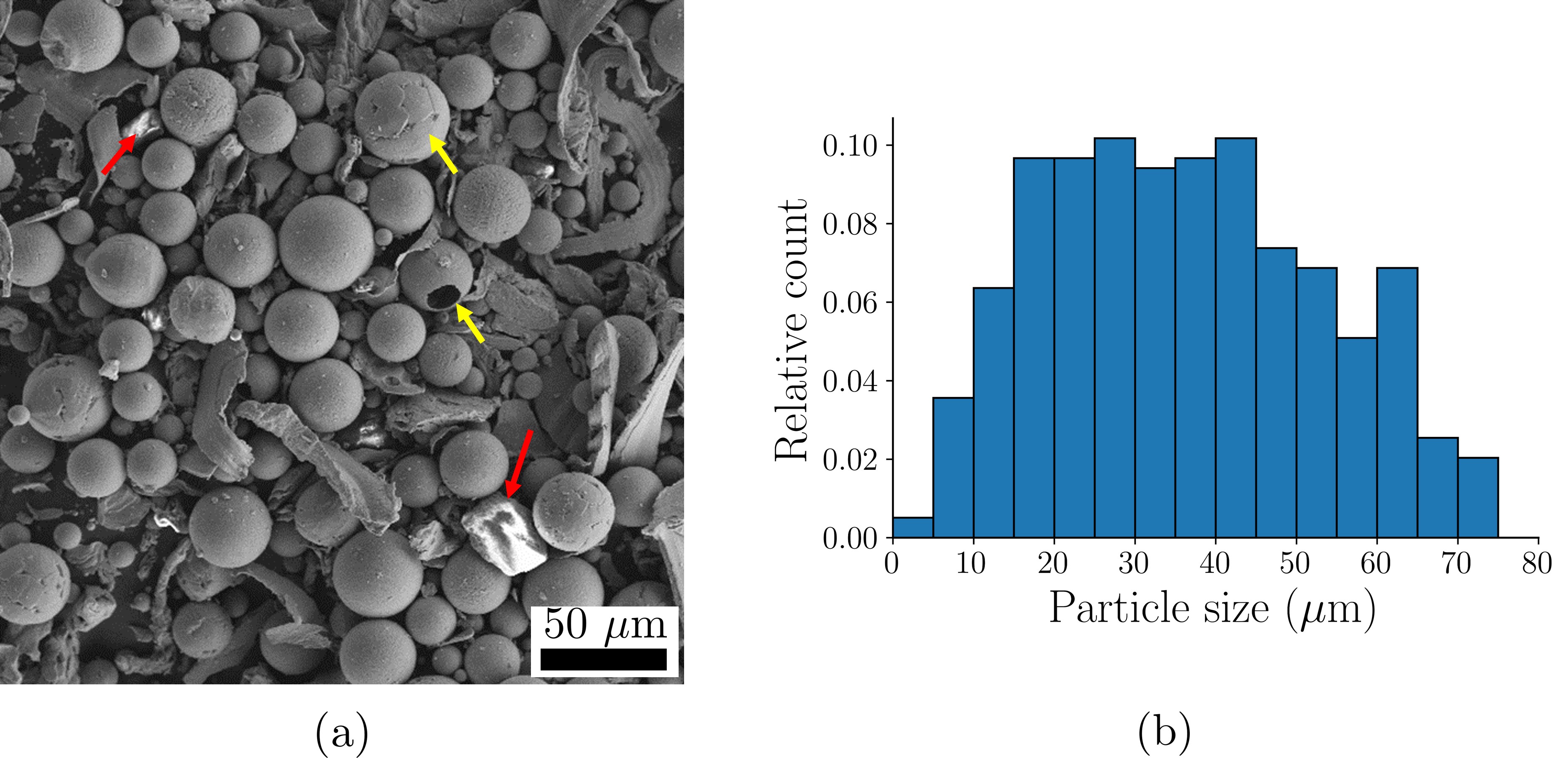}
  \caption{(a) SEM image of obtained spherical particles showing particles of perfectly spherical shape. Yellow arrows on image show potentially hollow spheres, red arrows show alumina particles. (b) Size distribution of spherical particles showing wide range of diamaters ranging from $10--70\,\mu$m.}
  \label{fig:SEM_images}
\end{figure}

The first figure, Fig.~\ref{fig:SEM_images}, discusses the size and shape of collected particles. Figure~\ref{fig:SEM_images}(a) is an SEM image of particles collected and sieved at a distance 300 mm from the wheel-workpiece contact zone. The particles are largely spherical in shape with stringy machining-type chips ($\sim 500-800\,\mu$m) also visible. The spherical particles occur in a variety of sizes, with some of them being hollow (at yellow arrows). A few abrasive particles (alumina) are also seen in the image (at red arrows) and were no doubt dislodged from the grinding wheel. Distribution of typical particle diameters is reproduced in Fig.~\ref{fig:SEM_images}(b). Clearly, particles with diameters between $10-70\,\mu$m appear to be predominant. The mean particle size is $\sim 30\,\mu$m and is smaller than mean abrasive size $\sim$ 150-200 $\mu$m in the grinding wheel.

\begin{figure}
  \centering
  \includegraphics[width=0.9\textwidth]{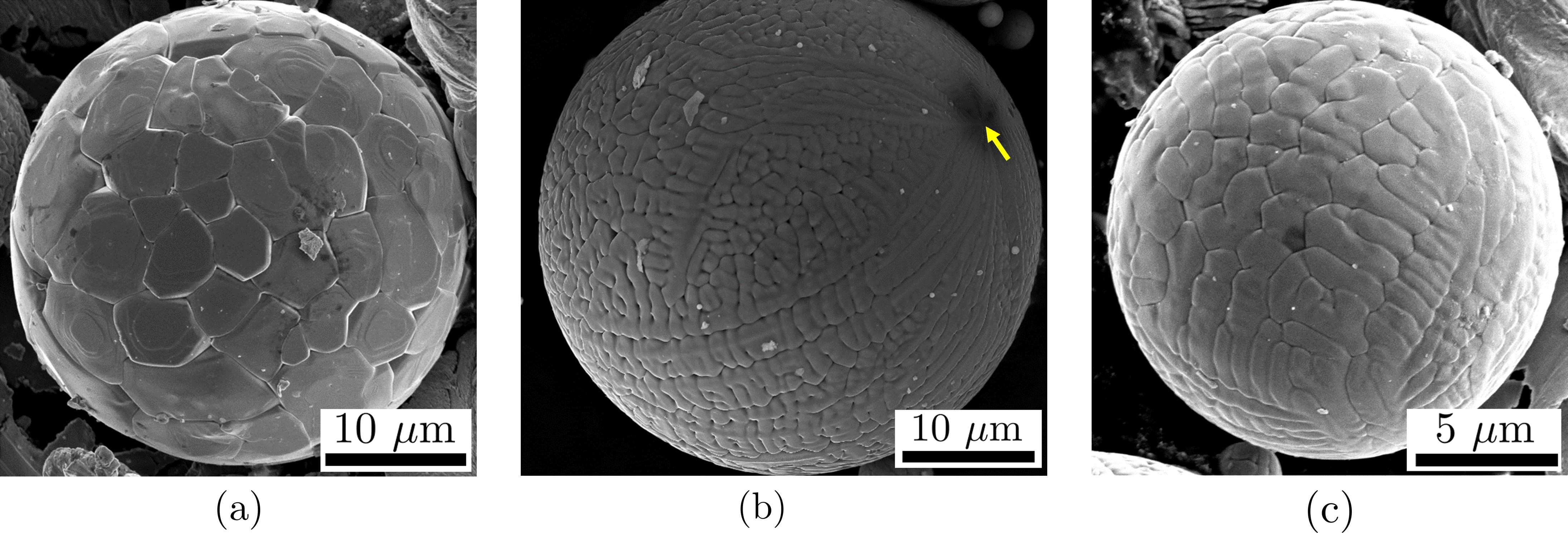}
  \caption{Scanning electron microsocpy images showing three distinct spherical particle morphologies. (a) Faceted structure resembling the surface of a soccer-ball. (b) Dendritic surface structure with arrow showing point of emergence of multiple primary branches. (c) Mixed surface morphology with broad dendrite arms.}
  \label{fig:SEM_morphologies}
\end{figure}

The second figure is comprised of three panels, each showing a spherical particle with distinct surface morphology, see Fig.~\ref{fig:SEM_morphologies}. Panel (a) shows a particle with typical cellular morphology and comprised of multiple facets on the surface. They are arranged together and resemble the patches on the surface of a soccer-ball. In panel (b) is a particle possessing dendritic surface morphology, with primary and secondary branches clearly visible. Several branches appear to emerge from a single point in the image (at arrow). Panel (c) shows an intermediate case between those of panels (a) and (b) and appears semi-dendritic with large branch spacing. All the particles collected were found to have one of these three morphologies. 

\begin{figure}
  \centering
  \includegraphics[width=0.6\textwidth]{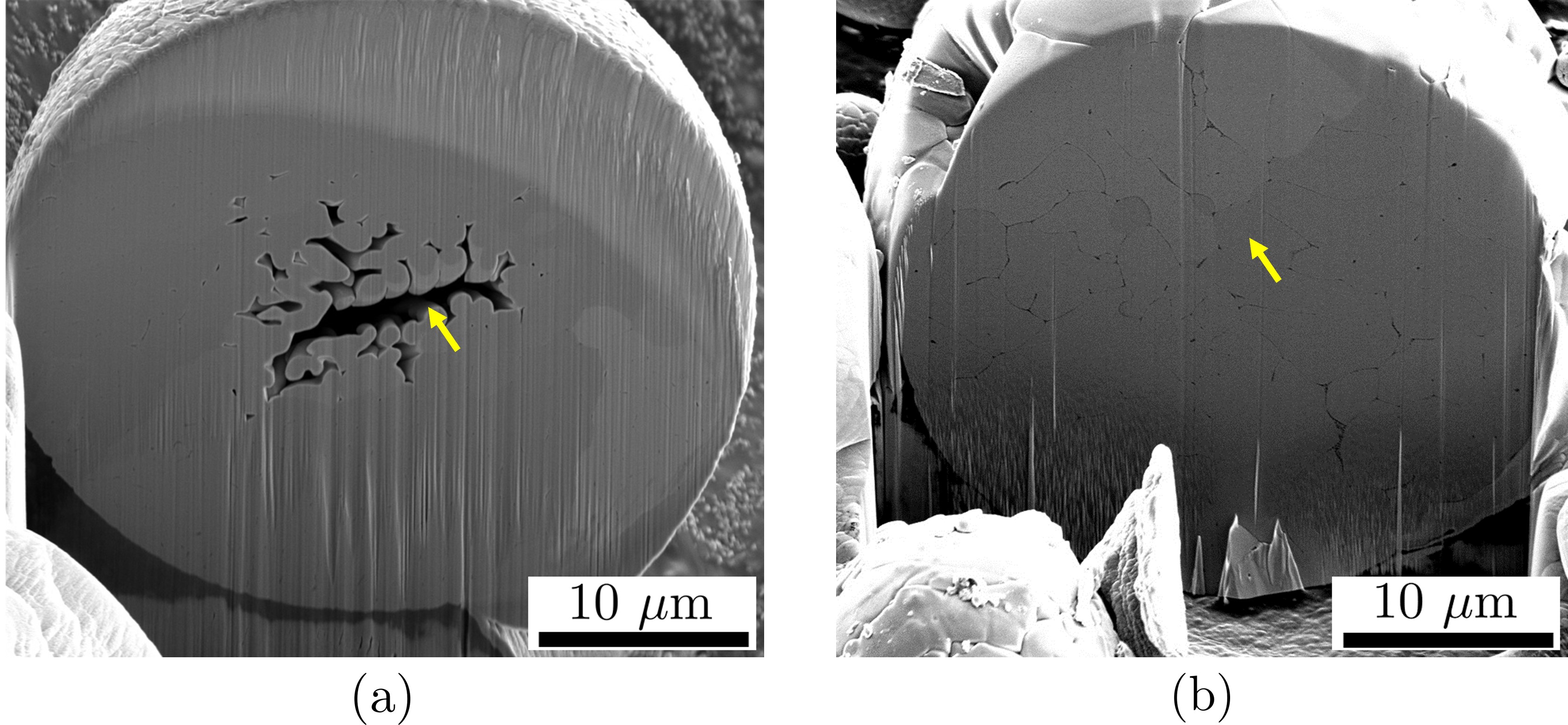}
  \caption{Scanning electron microscopy images of two typical cross sections of spherical particles. (a) Hollow core showing dendrite-like structures (at arrows) and (b) solid core with facet-like structures in the particle interior.}
  \label{fig:SEM_cross_sections}
\end{figure}

The third and final figure (Fig.~\ref{fig:SEM_cross_sections}) shows two cross sections typically observed in the resulting particles. The sections were obtained by FIB milling of single spherical particles (see Sec.~\ref{sec:experimental}). The image in Fig.~\ref{fig:SEM_cross_sections}(a) shows a particle with a hollow core (at arrow) and a branched inner structure, somewhat like the dendritic exterior seen in Fig.~\ref{fig:SEM_morphologies}(b). On the other hand, Fig.~\ref{fig:SEM_cross_sections}(b) shows a typical solid particle with a cellular-type inner structure (at arrow), analogous to the outer structure seen in Fig.~\ref{fig:SEM_morphologies}(a).

The near-perfect spherical shape, along with these morphological structures, are readily suggestive of solidification from a molten drop. In addition to these images, two other auxiliary observations are also noteworthy. Firstly, local temperatures in the grinding zone have never been recorded to exceed the melting point of steel \cite{hwang2003measurement, malkin2007thermal}. Consequently, the possibility of material melting within or near the contact zone purely due to interaction with the abrasive appears quite remote. Secondly, when performed in an inert environment (N$_2$ or Ar) and with the same steel sample, the number of spherical particles formed is nearly zero. Hence, it is clear that oxygen is necessary for the production of spherical particles in steel.

Based on these observations, we propose the following modified melting-resolidification mechanism for the development of a single spherical particle:
\begin{enumerate}
\item An abrasive causes material removal from the workpiece via cutting/rubbing action. The corresponding temperature rise during the abrasive--workpiece contact is large but not sufficient to cause local melting.
\item As the removed particle exits the contact zone, it oxidizes rapidly in ambient air. For Fe, this is a highly exothermic reaction that causes intense heat release, leading to melting of the particle.
\item The molten particle, made spherical by surface tension, then solidifies rapidly as it travels through the air. The morphologies in Fig.~\ref{fig:SEM_cross_sections} are a result of this rapid solidification process and likely involve multiple nucleation sites.
\end{enumerate}

Each of hypothesized steps (1)-(3) above are individually detailed in the next section (Sec.~\ref{sec:results}) using experimental evidence as well as complementary analytical calculations.

\section{Results and supporting evidence}
\label{sec:results}

We now present experimental evidence and corresponding analytical calculations that lend support to the hypothesis forwarded in Sec.~\ref{sec:hypothesis}. We start from the origin of a single particle---as a chip emerging from the workpiece---and follow its journey until it is collected.

\subsection{Temperature of removed material is lower than the melting point}
\label{subsec:thermalCalc}
The first stage is at the wheel-workpiece contact zone and involves interaction between abrasives on the wheel and a thin material layer. Determining whether or not melting occurs in this zone requires the evaluation of local or \lq flash\rq\ temperature due to the action of a single abrasive. Experimental measurement of flash temperatures in the contact zone is well-known to be nearly impossible in grinding, especially with the stochastic distributions of abrasive geometry and penetration depth \cite{malkin2007thermal}. This problem is further complicated by the near-complete absence of predictive theories for estimating temperature of the ejected chips---thermal heat partition problems in grinding almost always consider only the workpiece and wheel \cite{malkin2007thermal, guo1994analytical}. For the present case, prediction of the temperature rise in the ejected chip is crucial, while prediction of workpiece bulk temperatures can provide additional, supplementary information.

\subsubsection{Three-way heat partition in the contact zone}

\begin{figure}
   \centering
  \includegraphics[width=0.8\textwidth]{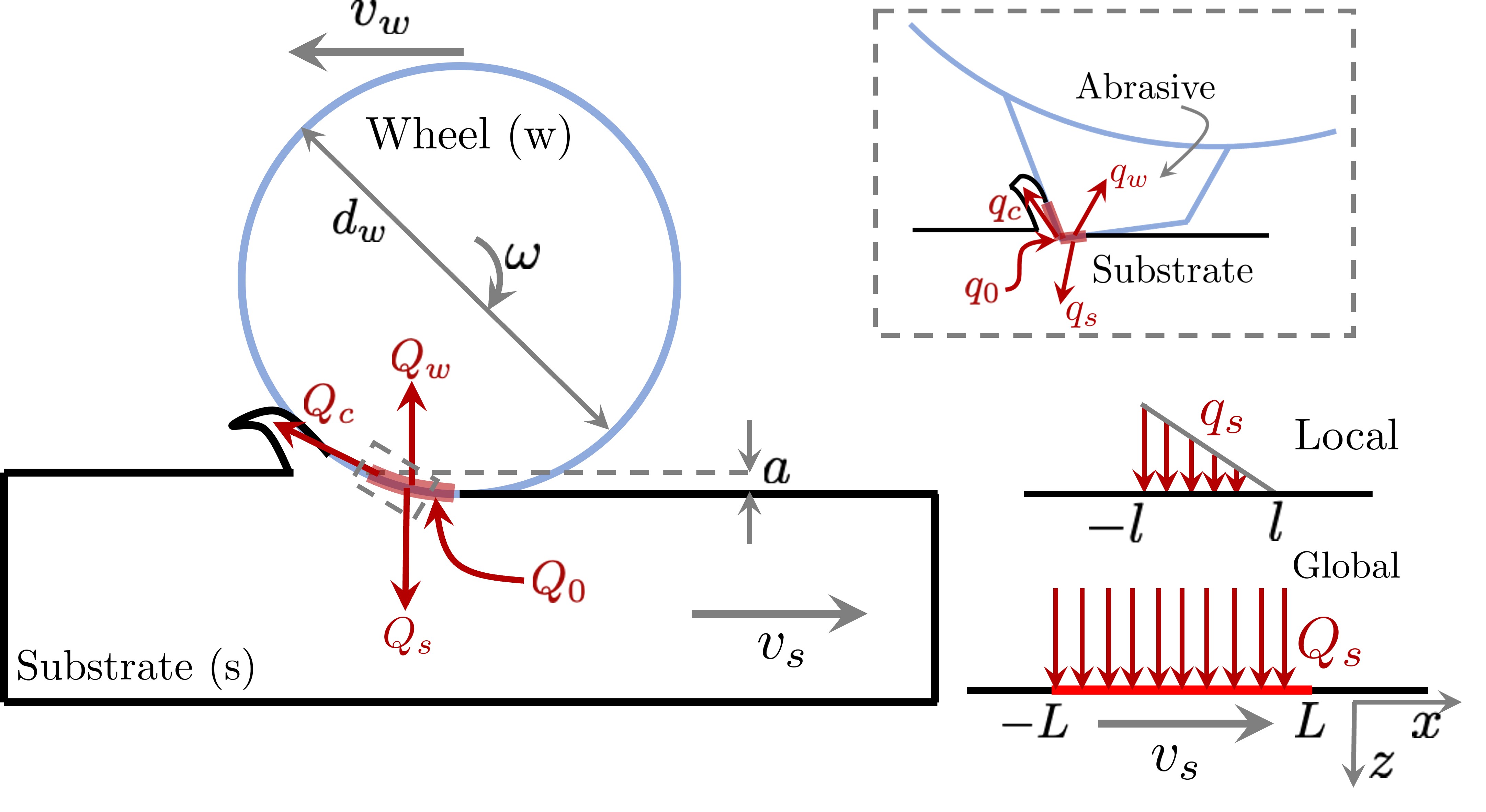}
  \caption{Schematic showing heat sources and heat partitioning during wheel-substrate interaction. On the global scale (left), heat $Q_0$ is generated at the contact, of which $Q_w, Q_c, Q_s$ go into the wheel, chip and substrate, respectively. On the local scale (see inset, dashed line), the heat at an individual asperity $q_0$ is partitioned into $q_s, q_w, q_c$ going into the substrate, wheel and chip, respectively. The model heat distributions for $q_s, Q_s$ are also shown (right).}
  \label{fig:grindingSchematic}
\end{figure}

To estimate the temperature rise in the chip as it exits the contact zone, we utilize the three-way heat partition model presented in our recent work \cite{DhamiJOM}, and originally proposed by Ju and colleagues \cite{ju1998theoretical}. Temperature in the chip and the workpiece are determined using coupled global-local heat partitioning, see schematic in Fig.~\ref{fig:grindingSchematic}. At the global scale, the wheel interacts with the workpiece, generating heat $Q_0$ in the process. Part of this heat goes into the wheel ($Q_w$), part into the substrate ($Q_s$) and the rest convected away by the removed material/ chips ($Q_c$). The key to determining the local temperatures is in evaluating how $Q_0$ is partitioned into $Q_w, Q_c, Q_s$. Using the fact that $Q_0 = Q_w + Q_c + Q_s$ from conservation of energy, we define heat partition coefficients $\epsilon_w, \epsilon_c, \epsilon_s$ as
\begin{equation}
  \label{eqn:partitionCoeff}
  \epsilon_w = Q_w/Q_0 \quad\quad \epsilon_c = Q_c/Q_0 \quad\quad \epsilon_s = 1 - \epsilon_c - \epsilon_w
\end{equation}
Assuming that $Q_0$ is known, the problem hence requires the determination of 2 unknowns $\epsilon_w, \epsilon_c$. Note that the global heat $Q_0$ is generated non-uniformly over the wheel-workpiece contact length $-L < x < L$, consequently each $\epsilon$ will be a spatially varying function. This will be explicitly accounted for later.

On the local scale, heat is actually generated via individual asperities interacting with the substrate. If the (known) ratio of nominal to real contact area is $\zeta$, then from energy balance, the local scale heat fluxes $q_0, q_w, q_c$ are related to the global values by
\begin{equation}
  \label{eqn:local_heats}
  q_0 = \zeta Q_0 \quad\quad q_w = \zeta Q_w \quad\quad q_c = \zeta Q_c
\end{equation}
These local heat sources are depicted in Fig.~\ref{fig:grindingSchematic} (inset). 

The local temperatures are then determined as follows, details of this scheme are presented in Ref.~\cite{DhamiJOM}. Firstly, suitable moving heat source models are used for each of the sources in Eq.~\ref{eqn:local_heats} and their global counterparts. All of these heat sources are spatially distributed so the choice of a model will depend on the corresponding contact geometry. For the global scale, we assume that the heat $Q_s$ is uniformly distributed along the contact length (Fig.~\ref{fig:grindingSchematic}(right)) and on the local scale, consistent with the abrasive cutting action, the heat $q_s$ is distributed in the form of a triangular source. We denote the local and global scale temperatures using Greek ($\tau$) and Latin ($T$) variables, respectively. Secondly, the temperatures are matched at each point $x$ of the corresponding interface according to the following scheme:
\begin{equation}
  \label{eqn:tempMatch}
  \tau_c(x) = \tau_w(x) \quad\quad \tau_w(x) = T_s(x) + \tau_s(x)
\end{equation}
The first relation implies that local abrasive and chip temperatures are equal at every point along their line of contact. The second one equates the total substrate temperature, comprising of both the local $\tau_s$ and global $T_s$ fields, is equal to the abrasive temperature at the local scale. This local equilibrium condition is well-established for intimate contacts as discussed in Ref.~\cite{DhamiJOM}.

In order to apply the conditions in Eq.~\ref{eqn:tempMatch}, we have to obtain the spatial temperature distribution along each of the contacting zones, as a function of the heat input (equivalently, $\epsilon_w, \epsilon_c, \epsilon_s$) using suitable temperature models. For the substrate on the global scale, the heat flux is assumed to be uniformly distributed in the form of a strip source of width equal to the contact zone width $L$. Consequently, the global substrate temperature is obtained as (see, for instance, Ref.~\cite{jaeger1942moving})
\begin{equation}
  \label{eqn:globalWPTemp}
  T_s(x)=\int_{-L}^{+L} \dfrac{\epsilon_s (x')Q_0(x')}{\pi K_s} \exp\left(\dfrac{V_s(x-x')}{2\alpha_s}\right)\times K_{0} \left(\dfrac{V_s|x-x'|}{2\alpha_s}\right)dx'
\end{equation}
where $v_s, \alpha_s, K_s$ are the substrate traverse speed, thermal diffusivity and thermal conductivity, respectively. $K_0(\rho)$ is the Bessel function of order 0 and $\epsilon_s$ is as defined in Eq.~\ref{eqn:partitionCoeff}. Note that this solution is obtained by solving the steady state heat equation for a continuously moving point source and using the principle of spatial superposition. 

For the local scale, the corresponding triangular source approximation (Fig.~\ref{fig:grindingSchematic}(right)) must be used. Now the steady state approximation cannot be employed since asperity contacts only exist for a short time duration, hence a full-time integral is employed as follows \cite{jaeger1942moving}:
\begin{equation}
  \label{eqn:localWPTemp}
  \begin{aligned}
\tau_s(x)=\int_{-L}^{+L} \dfrac{(1 - h(x^\prime))q_{s}(x')}{2 K_{s}}\sqrt{\dfrac{\alpha_{s}}{\pi v_{w} |x-x'|}} \erf \left(\dfrac{b_{g}}{\sqrt{\dfrac{4\alpha_{s}}{ v_{w} }|x-x'|}}\right)\times\\
\left[\erf \left(\dfrac{x-x'+l_{g}}{\sqrt{\dfrac{4\alpha_{s}}{ v_{w} }|x-x'|}}\right)-\erf \left(\dfrac{x-x'-l_{g}}{\sqrt{\dfrac{4\alpha_{s}}{ v_{w} }|x-x'|}}\right)\right]
\end{aligned}
\end{equation}
where $l_g, b_g, v_w$ are the single grain length, breadth and wheel linear velocity, respectively. The other constants are the same as earlier and the function $h(x) = 1$ if $x\geq 0$ and $h(x) = 0$ otherwise.

For the tool and workpiece, the following temperature models are used, as is common in the grinding literature \cite{DhamiJOM}. For the local abrasive temperature, a 1D approximation is used to obtain: 
\begin{equation}{\label{eqn:grainTemp}}
 \tau_{t}(x)=\tau_{t0}+\int_{-L}^{+L} \dfrac{h(x)q_{t}(x')}{ K_{t}}\sqrt{\dfrac{\alpha_{w}}{\pi v_{w} |x-x'|}}dx'
\end{equation}
Where, $\tau_{t0}$ is the initial grain temperature. Here again, the coefficients $K_t, \alpha_w$ correspond to the grain thermal conductivity and diffusivity, respectively. For the chip temperature, the heat flux is approximated as being triangular so that the final temperature distribution is
\begin{equation}{\label{eqn:chipTemp}}
  T_{c}(x)=\epsilon_c(x) \dfrac{q_{c}(x)}{A \rho_{s} C_{s}}\dfrac{2 L^2}{\gamma (L+x)}
\end{equation}
Where, $\rho_{s}, C_s$ are mass density and specific heat of the chip/substrate material, $\gamma$ is the imposed material removal rate per unit width.

Finally, we can use the temperature fields Eq.~\ref{eqn:globalWPTemp}--\ref{eqn:chipTemp} in the two matching conditions of Eq.~\ref{eqn:tempMatch} to obtain expressions for the partition coefficients $\epsilon_c, \epsilon_w, \epsilon_s$. The end result are two equations, valid point-wise along the contact that must be solved numerically. Details of this calculation, including the discretization step and a typical solution scheme using Simpson's 1/3 rule are provided in Ref.~\cite{DhamiJOM}.

\subsubsection{Lower-bound temperature estimate}
The three-way heat partition described above can be used to obtain a lower bound temperature for the chip as it exits the contact zone. For this purpose, we have to solve for the partition functions in Eq.~\ref{eqn:partitionCoeff} using the relations in Eq.~\ref{eqn:tempMatch}. In order to obtain values for the experimental configuration described in Sec.~\ref{sec:experimental}, the following parameters were used: For the substrate (AISI 52100 steel), $K=60.5$ W/m-K, $\alpha= 1.8 \times 10^{-5}$ m$^2$/s. For the abrasive (alumina), $K=36$ W/m-K, $\alpha= 9.8 \times 10^{-6}$ m$^2$/s. Geometric parameters were nominally set at $Q_0=88$ W/mm$^2$, $A=0.25\%, 2l_g=2b_g=0.25$ mm, $L=1.25$ mm. These are representative values for the grinding wheel used in the experiments.

\begin{figure}
   \centering
  \includegraphics[width=0.8\textwidth]{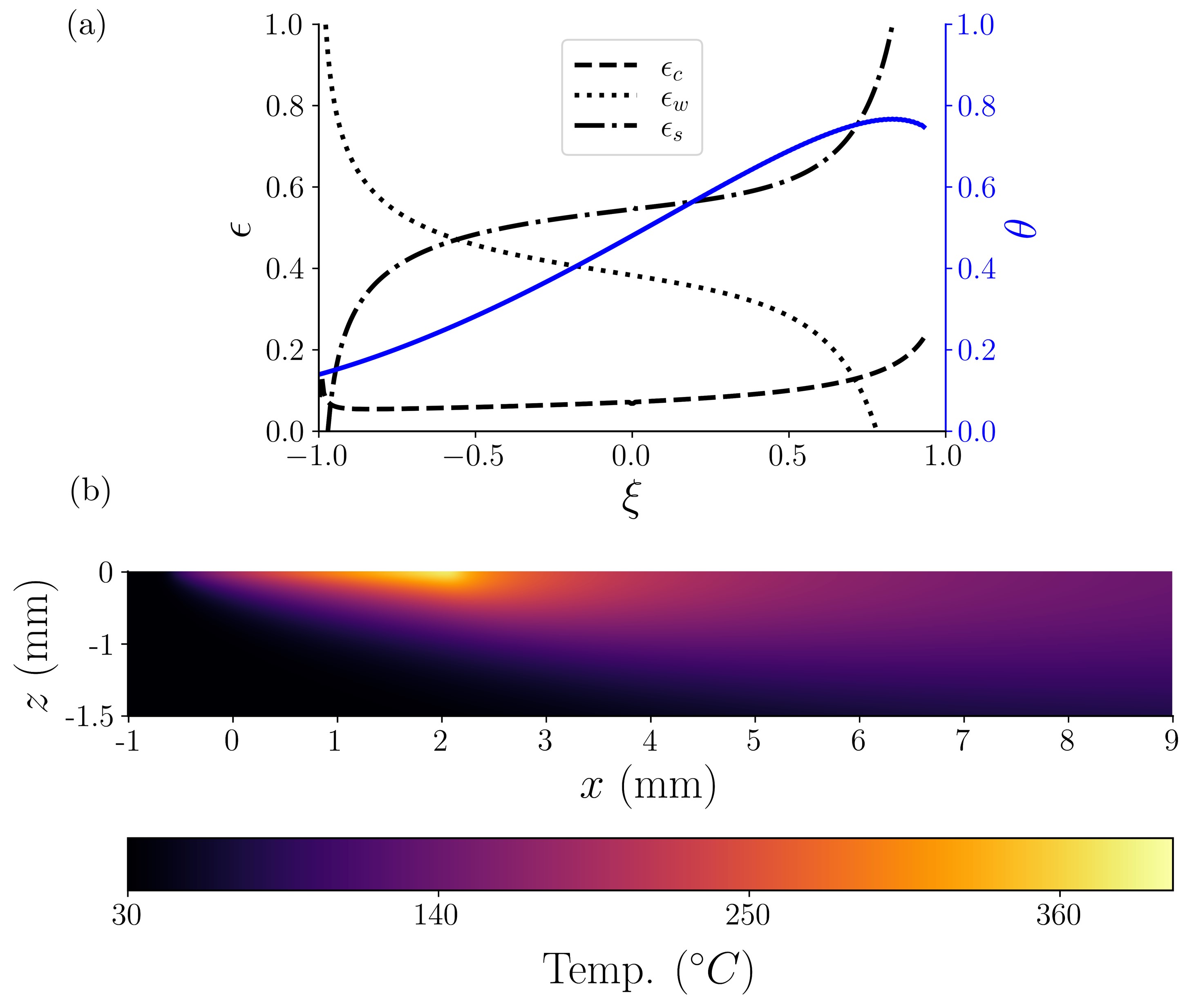}
  \caption{Analytical temperatures obtained using the three-way heat partitioning scheme. (a) Partition functions $\epsilon_c, \epsilon_w, \epsilon_s$ along the corresponding non-dimensionalized contact length $\xi$. (b) 2D temperature field showing variation into the depth of the substrate ($z$-direction). The maximum substrate temperature of $\sim 400^\circ$C occurs at the substrate surface. }
  \label{fig:tempFields}
\end{figure}

The resulting partition functions are shown in Fig.~\ref{fig:tempFields}(a) as a function of the non-dimensional distance $\xi = x/L$ within the contact zone. It is clear from this figure that, for the present experimental configuration, we should expect the substrate to take away most of the heat within the contact zone, see $\epsilon_s$ curve (dash-dot line) in Fig.~\ref{fig:tempFields}(a). In contrast, the fraction of heat carried away by the chip is quite small, $\epsilon_c \sim 6\%$ on average. This is not surprising, and is the reason that nearly all analysis of grinding temperatures neglect heat lost to the chip. These functions can then be used to evaluate the temperatures of the wheel, chip and substrate, when coupled with the corresponding temperature fields which they parametrize. 

Given the time and length scales involved, it is clear that the chip temperature is not directly experimentally measurable using IR thermography. Instead, we obtain a lower bound on the chip temperature by measuring the workpiece temperature rise and comparing it with the predictions of the three-way partition model. The corresponding spatial non-dimensionalized substrate temperature within the contact zone, defined as $\theta_s = K_s (T_s + \tau_s)/(Q_0 L)$ is shown in Fig.~\ref{fig:tempFields}(a), see blue solid curve. The temperature increases from the engagement point ($\xi = -1$) to the disengagement point ($\xi = +1$) with the wheel. The maximum temperature occurs inside the contact zone, closer to the end $\xi = +1$, it approaches this end as the velocity of the wheel is further increased.

The accompanying 2D temperature field into the depth of the steel substrate ($-z$ direction) is shown in Fig.~\ref{fig:tempFields}(b). Here the wheel contact zone is between 0--2.5 mm and the maximum temperature rise occurs on the surface within the contact zone, as described earlier in Fig.~\ref{fig:tempFields}(a). It is clear that the model computations negate the possibility of melting occurring on the workpiece surface since the maximum predicted temperature is $\sim 400^\circ$C. Likewise, the corresponding temperature rise in the chip (Eq.~\ref{eqn:chipTemp}) is also predicted to be much lesser ($\sim 650^\circ$C) than the melting point ($\sim 1400^\circ$C).

\begin{figure}
   \centering
  \includegraphics[width=\textwidth]{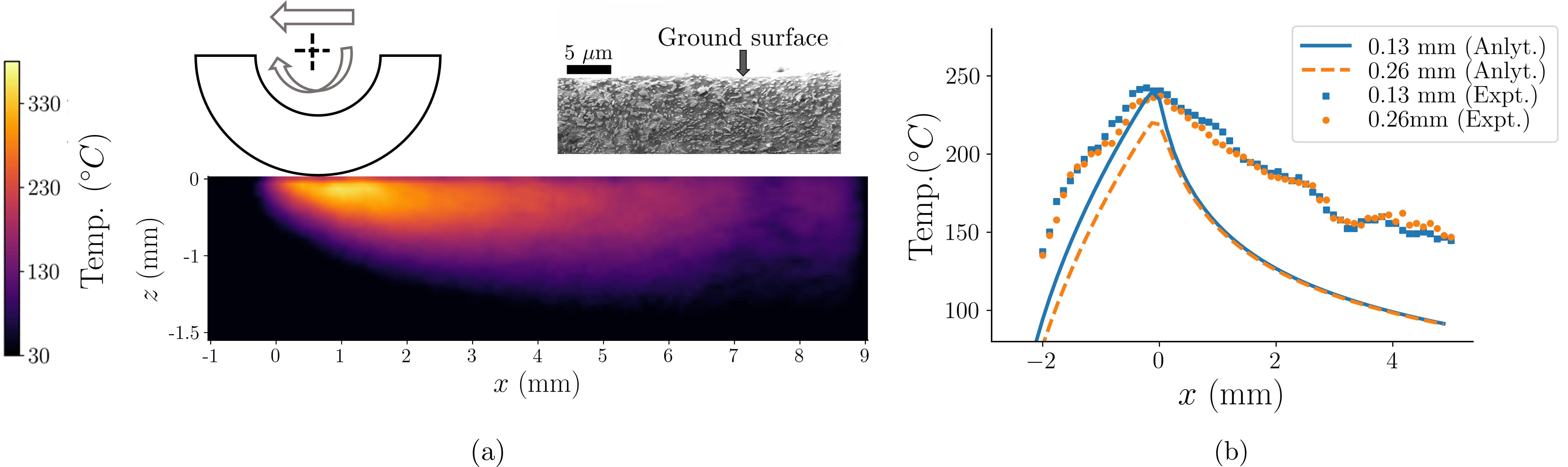}
  \caption{Experimentally observed substrate temperature fields using IR thermography. (a) 2D temperature distribution ($xz$-plane) within the workpiece appears to closely match the predicted fields from the three-way partition analysis. Inset to this figure shows an image of the $xz$ section of the workpiece. No sign of any recrystallization or grinding burns are visible. (b) Predicted sub-surface temperature fields match the measured fields quite closely. The maximum sub-surface temperature in the substrate does not exceed $400^\circ$C. }
  \label{fig:tempFieldExpt}
\end{figure}

These model predictions for the substrate temperature were compared with experimental measurements obtained using infrared thermography, and also evaluated the substrate surface condition post grinding, see Fig.~\ref{fig:tempFieldExpt}. The thermal field in Fig.~\ref{fig:tempFieldExpt} must be compared with that predicted by the theory (\emph{cf.} Fig.~\ref{fig:tempFields}). Despite a minor difference in the maximum predicted temperature ($400^\circ$C vs. $\sim 360^\circ$C), the spatial distribution in the two cases appears to be nearly identical. Furthermore, microscopic observations of the cross section of the ground surface (inset to figure) show no signs of any consequent phase transformations (e.g., martensite formation), consistent with these temperature predictions. It is thus safe to conclude that the workpiece does not even reach its recrystallization temperature, let alone its melting point.

\subsubsection{Upper-bound temperature estimate}

The analysis above predicts larger (lower) temperatures for the substrate (chip) and therefore constitutes a lower bound estimate for particles ejected from the contact zone. While the chip temperature is not expected to differ significantly from the values obtained here, it is nonetheless worthwhile to obtain an upper bound estimate for the chip temperature to completely rule out the possibility of melting.   

In order to do this, we simply neglect global-scale heat loss to the bulk of the substrate and the wheel and evaluate partition of heat only between the abrasive and the chip. This is most easily done by viewing this as a tool-chip contact, typical of a machining process, with the caveat that the tool (abrasive) has a large negative rake angle. We are then faced with a simpler partition problem involving only two objects and one partition function $\epsilon_0$ \cite{loewen1954analysis}. There are, however, two sources of heat, one from the primary shear zone (between chip and substrate) and the other at the secondary shear zone (abrasive/chip). Consequently, the total temperature rise (non-dimensionalized as before) in the chip can be obtained as:
\begin{equation}
  \label{eqn:chipTempUB}
  \theta_c =\dfrac{1.13(1-\epsilon_0)q_1 l_0}{K_c}\left[\dfrac{1}{\sqrt{Pe}}\right]+\dfrac{\zeta q_0 l_1}{K_t}\left[\dfrac{1}{1+\dfrac{\zeta\sqrt{Pe_1}}{1.13}}\right]
\end{equation}
where the non-dimensional parameters $Pe, Pe_1$ represent the P\'eclet numbers corresponding to the primary and secondary shear zones, respectively. The P\'eclet number $Pe = uL/\alpha_c$ is obtained using a velocity scale (either shear or chip velocity), a length scale (shear plane length or tool-chip contact length), and chip thermal diffusivity $\alpha_c$. Here, $\zeta$ is termed the area factor \cite{loewen1954analysis}, $l_0$ is the length of the shear plane, $K_c$ and $K_g$ are thermal conductivity of the chip and abrasive, respectively. Finally, $q_0, q_1$ are the heats generated at the primary and secondary shear zones, respectively, and depend on the abrasive rake angle $\alpha$ and the depth of cut $t_0$. These are evaluated as follows: $q_0=\dfrac{F_s V_s}{A_p}$, and $q_1=\dfrac{F_f\times V_c}{A_s}$ where, $F_s, F_f$ are the shear and friction forces, $V_c, V_s$ are cutting and shear-plane velocities and $A_p, A_s$ are the area of primary and secondary shear zones, respectively. Using the Merchant model for forces, along with the minimum energy criterion, we obtain the final temperature as a function of the kinematic parameters and shear strength $\tau_0$ of the material \cite{merchant1945mechanics}. 

\begin{figure*}[h!]
   \centering
  \includegraphics[width=0.6\textwidth]{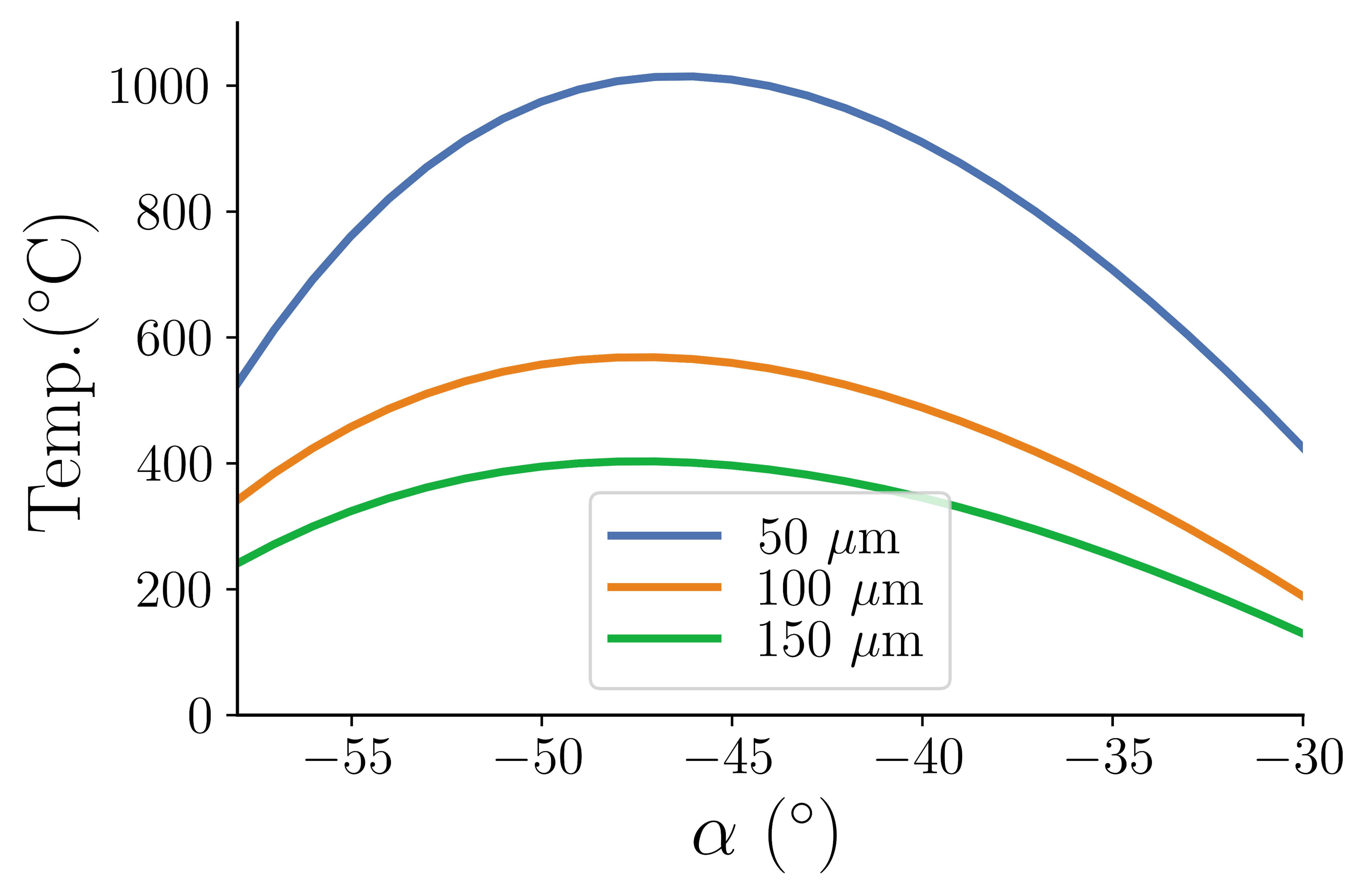}
  \caption{Upper bound estimates of chip temperatures for three different depths of cut of the abrasive. The curves show that a maximum of 1000$^\circ$C is possible only at the extremely unlikely cutting geometry of $\alpha = -47^\circ$.}
  \label{fig:chipTempUB}
\end{figure*}

The expression in Eq.~\ref{eqn:chipTempUB} is for the total temperature rise in the chip and can now be evaluated for a range of grain-substrate interaction parameters---the depth of cut and the rake angle---which will change $q_0, q_1$. For the present analysis, values of $\tau_0$, $F_c$, $F_s$ and $F_f$ have been calculated based on the nominal input parameters for high speed cutting operation which are adapted from Ref.\cite{Shaw2005}. The area factor $\zeta = 1.4$ 1.4 and $V_c= 10$ m/s. Standard values of thermal properties of the work piece and tool have been used. The result is shown in Fig.~\ref{fig:chipTempUB}. The three curves in this figure correspond to three different depths of cut ranging from 50--100 $\mu$m. It is evident from the figure that the largest chip temperature $\sim 1000^\circ$C is obtained with a shallow $50\,\mu$m depth of cut and rake angle $\alpha = -47^\circ$. This clearly provides an upper bound estimate for the chip temperature since abrasive grain rake angles are seldom larger (less negative) than $-65^\circ$ to $-70^\circ$ \cite{malkin2007thermal}. Under these conditions, the temperature rise is again not significantly larger than the $\sim 650^\circ$C rise predicted using the three-way partition method. Hence, it is quite safe to conclude that chip melting is highly unlikely to occur (if at all) as material is ejected from the contact zone.

\subsection{Exothermic surface oxidation causes particle melting}
In line with the proposed hypothesis (Sec.~\ref{sec:hypothesis}), the next stage of the ejected particle's life is traversal through air. In the presence of ambient oxygen, the particle is postulated to undergo rapid oxidation (exothermic reaction), leading to significant energy release. This energy is sufficient (as is shown below) to raise the total particle temperature to well above the melting point.

\begin{figure*}[h!]
   \centering
  \includegraphics[width=0.75\textwidth]{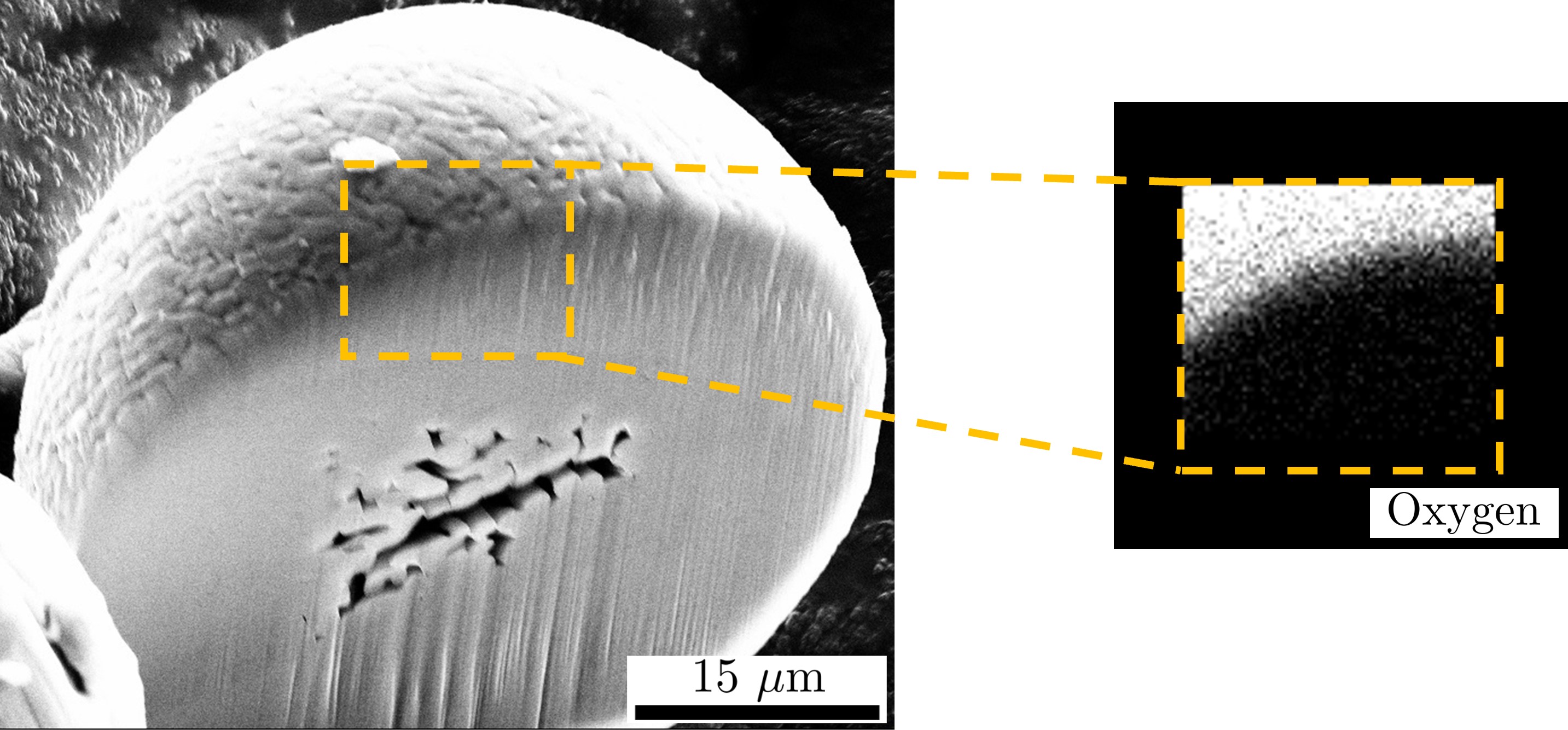}
  \caption{Elemental map showing spatial distribution of elemental oxygen within a thin layer of thickness $\delta$ around $1\%$ of the sphere's radius.}
  \label{fig:oxidation}
\end{figure*}

We first present experimental evidence for the occurrence of oxidation using \emph{post mortem} analysis of the final spherical particles. A single spherical particle was first sectioned using FIB milling and an EDS spectrum extracted from the cross section, see Fig.~\ref{fig:oxidation}. The presence of a thin oxide layer is clear from this data, with insignificant or complete absence of oxygen away from the particle's periphery. Typical weight percentages of oxygen ($41.3\%$) and iron ($58.7\%$) within this layer are adequate for the formation of iron oxide Fe$_3$O$_4$. Further, the layer thickness is not more than $1-2\,\mu$m, indicating that only a thin surface layer of the ejected particle is subject to oxidation. Traces of diffused oxygen were also registered inside the sphere but their concentration was found to be insignificant in comparison to the surface.

\begin{figure*}[h!]
   \centering
  \includegraphics[width=0.8\textwidth]{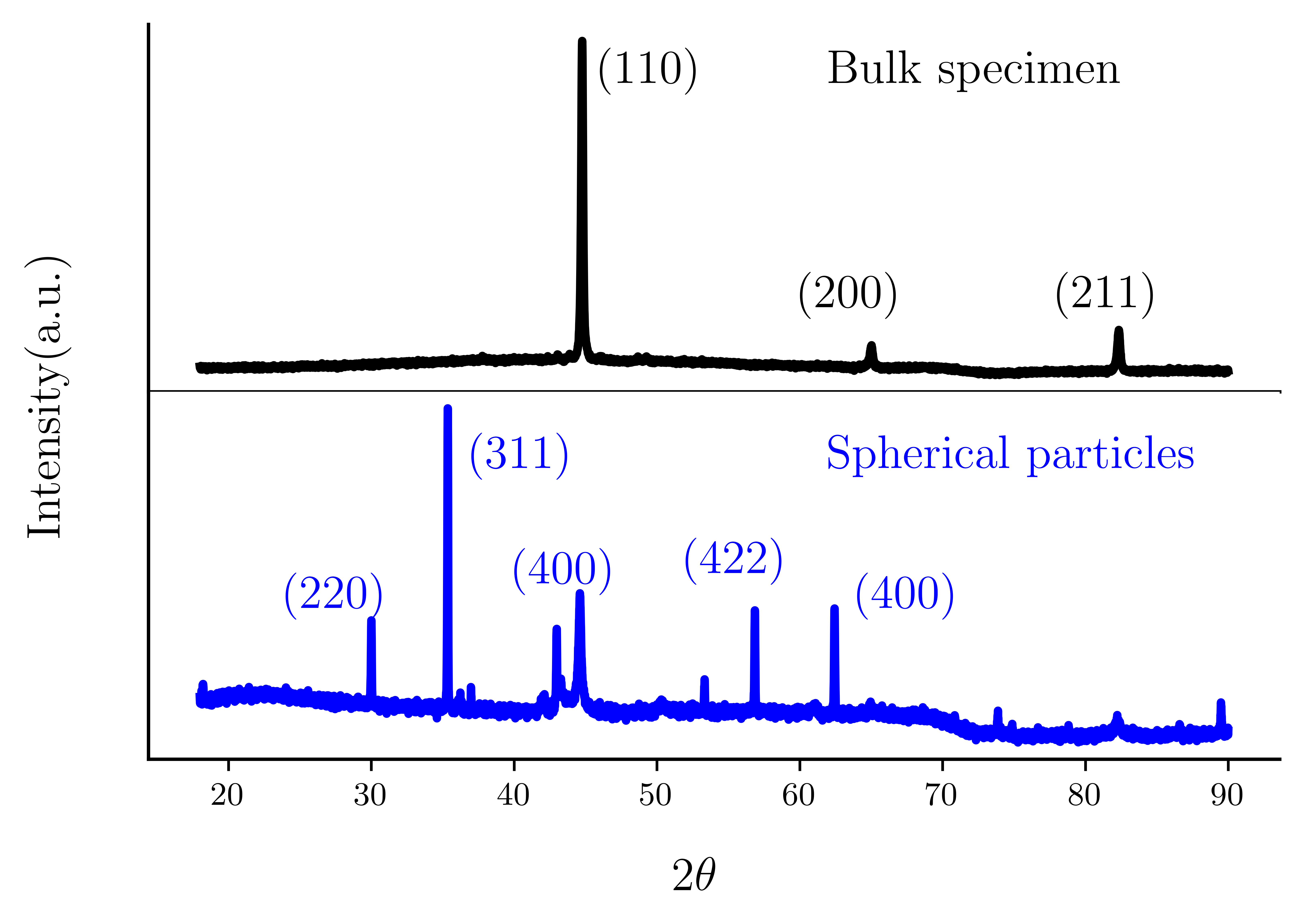}
  \caption{X-ray diffraction data for the bulk specimen (top row) and the collection of spherical particles (bottom row). Peaks corresponding to Fe$_3$O$_4$ are clearly visible in the bottom plot and are absent in the top.  }
  \label{fig:XRD}
\end{figure*}

X-ray diffraction analysis helped shed further light on the oxidation route, see Fig.~\ref{fig:XRD} showing diffraction data for the bulk specimen (top) and spherical particles (bottom). The peaks were matched with ICDD/JCPDS data, XRD of the bulk specimen (before grinding) shows diffraction peaks at 44.9, 65.7, and 83.3 degrees, which correspond to standard peaks of $\alpha$-Fe BCC phase (110),(200), and, (211) respectively. On the other hand, XRD spectrum of grinding swarf has additional peaks at 29.6, 34.7, 36.47, 54.1, 59.8 and, 61.4 degrees, which can be attributed to Fe$_3$O$_4$ (220),(311), (400), (333), (422) and (440) respectively. These peaks were matched with ICDD data for Fe$_3$O$_4$ file no JCPDS-19-629.

In order to explain these observations, as well as to estimate both the oxide layer thickness and the oxidation time, we begin with a standard parabolic oxidation law \cite{pilling1923oxidation, cabrera1949theory}
\begin{equation}
  \label{eqn:parabolic}
  X^2=K_{p} t +C
\end{equation}
where $X$ is the mass of oxygen consumed per unit surface area, $K_{p}$ is a rate constant (determined experimentally) and $t$ is the time taken for oxidation. At elevated temperatures, $K_p$ obeys an Arrhenius-type relation \cite{campbell1947oxidation}. For alloy steels, we use data for oxidation of iron in an oxygen atmosphere \cite{smeltzer1960kinetics, davies1951mt} to obtain $K_P \sim 5\times 10^{-10}$ g$^2$/(cm$^4$ sec) for a particle ejected from the grinding contact zone. It must be noted that the parabolic law breaks down as we approach the melting point \cite{moulin1980influence} so that in the present case, where the chip exit temperature is much lower than the melting point, our assumption appears reasonable. Using Eq.~\ref{eqn:parabolic}, for the mass of a single particle, we obtain an upper bound on the oxidation time $t_{ox} \sim 10^{-6}$ seconds. Naturally, if only a small spherical shell were subject to oxidation, this time would be significantly lower. Either way, the oxidation time is much smaller than the travel time of the particles from the grinding zone to the collection bin so that the oxidation reaction may be assumed to occur practically instantaneously.

Consequently, the exothermic oxidation reaction acts as an instantaneous source of heat within the particle that does not have sufficient time to be convected away. If the thickness of the oxidized layer is $\delta$, the corresponding Biot number for the drop under convective cooling is $Bi \sim 10^{-3}$, so that the entire particle may be treated as a lumped object. Using simple energy balance, the temperature rise $\Delta T$ may be estimated from the oxidation energy release $\Delta H$ as:
\begin{equation}{\label{eqn:oxideThickness}}
	\frac{\delta}{R_0}  = \frac{L+C_{c} \Delta T}{\Delta H} 
\end{equation}  
where $R_0$ is an effective size measure of the particle and $L, C_c$ are the latent heat of fusion and specific heat, respectively. This energy balance (Eq.~\ref{eqn:oxideThickness}) may be used to provide an upper bound on the oxide layer thickness if the heat released $\Delta H$ were sufficient to melt the entire particle. For this, we use $\Delta T = 900^\circ$C to equal the difference between the particle's exit temperature (Sec.~\ref{subsec:thermalCalc}) and its melting point. The value $\Delta H = -265$ kJ/mol is chosen for Fe $\to$ Fe$_3$O$_4$ (\emph{cf.} Fig.~\ref{fig:XRD}). Using these values, we obtain
\begin{equation}
  \delta/R_0 \simeq 0.01
\end{equation}
For a spherical particle with $R_0\simeq 50\,\mu$m, a corresponding oxide layer of thickness $\delta = 0.5\,\mu$m would be sufficient to melt the entire particle, \emph{cf.} Fig.~\ref{fig:oxidation}. Once an oxide layer is formed on the surface, it acts as a barrier for the diffusion of oxygen towards the center of the particle so that subsequent oxidation is significantly delayed. These results establish that it is the oxidation stage that is likely responsible for the melting of the particle and, consequently, its perfectly spherical shape. Further, it is almost certain that this oxidation process is responsible for the spark/ streak observed just beyond the contact zone.

\subsection{Rapid solidification results in characteristic microstructures}
\label{subsec:solidification}

Post-oxidation and melting, the ejected particle is now in the form of a spherical molten drop as it approaches the collection region. Consequently, it undergoes convective cooling and solidification over a timescale of $10-100$ milliseconds to form a spherical particle. This constitutes the final stage in its evolution; the associated rapid solidification process leaves behind its own set of signatures that we now discuss.

Recall that three distinct microstructures are observed in the powder particles---cellular, dendritic and mixed, \emph{cf.} Fig.~\ref{fig:SEM_morphologies}. The cooling rate is the key parameter that determines which of these structures actually occurs in a drop of a given size. In the present problem, the cooling rate for the drop to form a powder particle is $\sim 10^5\,^\circ$C/s. At these high solidification rates, high carbon steel is likely to form body centered tetragonal (BCT) martensitic structure\cite{krauss1999martensite}. Evidence for this is not discernible from the XRD analysis of Fig.~\ref{fig:XRD}) since both BCC and BCT structures show practically overlapping peaks at  $2\theta$=44.7$^\circ$ in the XRD spectrum \cite{chen2018cubic}.

\begin{figure}[h]
\centering
\includegraphics[width=0.8\linewidth]{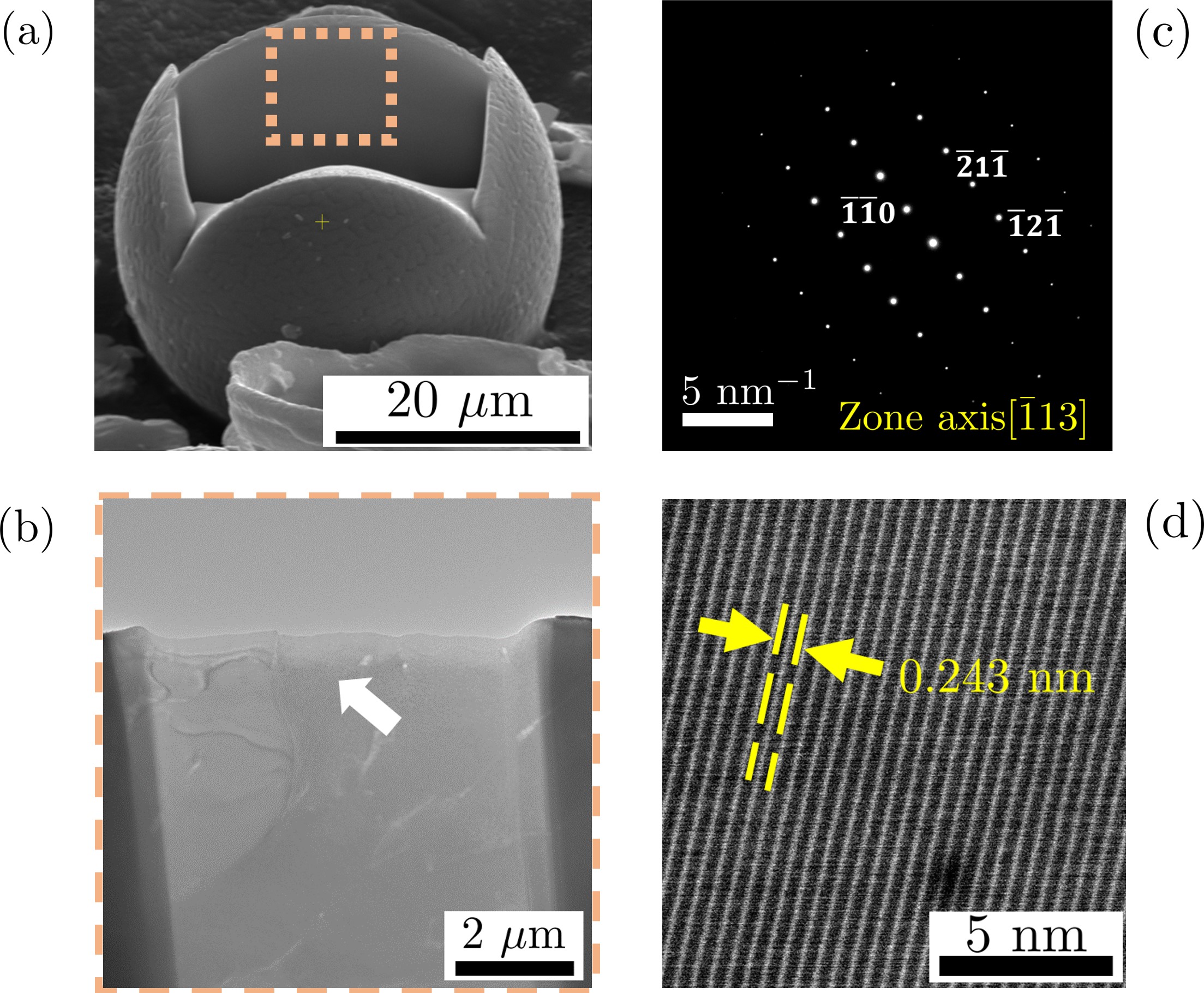}  
\caption{Transmission electron microscopy of a single particle reveals lattice structure of solidified Fe. (a) SEM image showing cross section where sample extraction was performed. (b) Thinned sample showing location of TEM data collection site (at arrow). (c) Selected area diffraction (SAD) pattern showing the various spots corresponding to lattice directions for both BCC and BCT Fe. (d) HRTEM image reveals a corresponding spacing of 2.43 \AA .}
\label{fig:TEM}
\end{figure}

In order to investigate the structure of the crystal lattice, we performed TEM analysis on a sample section retrieved from the cross section of a typical spherical particle, see Fig.~\ref{fig:TEM}(a). The particle cross sections were obtained in a manner similar to that in Figs.~\ref{fig:SEM_cross_sections} and~\ref{fig:oxidation} using FIB milling. TEM selected area diffraction (SAD) and high resolution-TEM (HRTEM) data were taken at different distances from the outer surface of the particle and towards the center, see Fig.~\ref{fig:TEM}. The obtained SAD patterns were indexed using zone axis [$\bar{1}13$]. The corresponding lattice parameters were calculated from SAED patterns to be $a=2.54\AA,  c=2.58\AA$, which give $c/a$ ratio of 1.02. These values are consistent with lattice parameters of BCT martensitic phase in high carbon steel \cite{lu2017effect}. Additionally, HRTEM analysis of the sample showed the $d$-spacing of 0.243 nm. These results, taken together, appear to strongly point at the occurrence of iron in the BCT martensitic phase, as is to be expected from the rapid solidification hypothesis.

The question now remains of explaining why each of the solidification morphologies of Fig.~\ref{fig:SEM_morphologies} occur. Contrary to conventional diffusion-controlled solidification processes, the conditions under which planar--cellular--dendrite morphology transitions occur during rapid solidification remain an area of active investigation \cite{galenko1997local}. Even the simplest analysis of a perfectly stable spherical solidification front presents significant difficulty \cite{DavisHill1982, PrudhommeETAL1989}. The primary complication arises from the unknown location of the solid--liquid interface during solidification, which must be determined as part of the solution---the so-called Stefan problem in conduction heat transfer. A secondary complication is the singular nature of the geometry. Unlike conventional slab casting, a drop has a finite volume and if solidification proceeds inwards from the outer surface, the solution becomes singular as the front approaches the drop center \cite{mccue2008classical}. Consequently, even a stable interface growth problem remains challenging to solve analytically; potential interface instabilities leading to cellular/dendritic growth remain, at present, intractable except by very approximate methods \cite{gill1981rapid, gill1984rapid}.

A third, more serious, complication and one that has not even been considered by the works cited above is the possibility of multiple nucleation sites. Under rapid solidification, alloys, such as the steel used here, inevitably solidify by the formation of several nuclei on the surface \cite{levi1982heat, Markworth1988}. In fact, additional images of similar morphologies suggests the cells in Fig.~\ref{fig:SEM_morphologies} actually constitute individual nucleation sites as opposed surface features arising from a growing unstable spherical front. Given the seemingly insurmountable complexities of the thermal field surrounding multiple nuclei, we resort to a kinematic model to explain some of the void features in the cellular patterns.

\begin{figure}[h]
\centering
\includegraphics[width=0.8\linewidth]{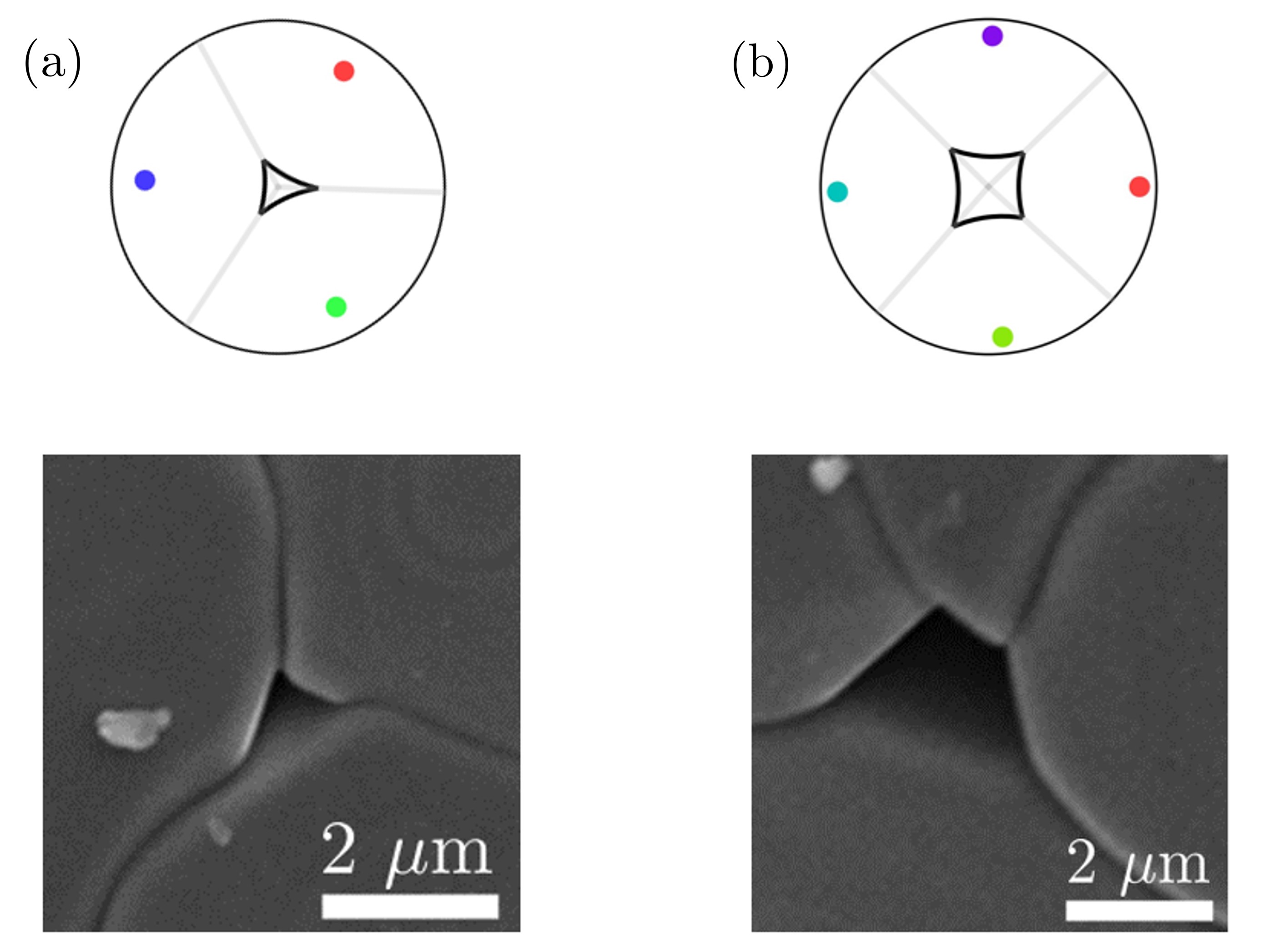}  
\caption{Kinematic nucleation model for describing voids showing (a) 3-site (top numerical, bottom expt.) and (b) 4-site solutions (top numerical, bottom expt.)}
\label{fig:solidification}
\end{figure}

We begin by considering a 2D circle with multiple nucleation centers located on its periphery---generalization to the 3D case is straightforward and is discussed below, see Fig.~\ref{fig:solidification}. Consider a 2D circular drop of radius $R_0$ with $n$ nucleation sites along its circumference. On the microscopical ($1\,\mu$m or more), crystals centered at each of these points grow with circular symmetry and first meet at a point midway between the two nucleation centers, termed $N_1, N_2$ in the figure. Further growth proceeds along the perpendicular bisector of $N_1 N_2$---if the growth rate ratio between the two crystals is $\xi$, then this line of intersection of the two crystals will be distant $\xi$ from one of the nuclei, yet still perpendicular to $N_1 N_2$. Taking the final crystal radius to be $r$, with $r<R_0$ being distinct from the drop radius due to solidification shrinkage, then the shape of the void is determined using elementary geometry as shown in Fig.~\ref{fig:solidification}.

The only changes to this calculation for the case of a 3D sphere with arbitrary nucleation sites are as follows. Firstly, since growth is now spherically symmetric, the line of intersection between adjacent nuclei will become instead a plane passing through the sphere center and perpendicular to the line $N_1 N_2$. The intersection of this line with the drop surface will be a great circle. Consequently, the original line $AA^\prime$ in Fig.~\ref{fig:solidification}(a) must be interpreted as the planar projection of a circle on the surface. Secondly, if the nucleation points are now moved closer to each other on the surface of the sphere, the calculation remains the same with the final void shape being determined only by $r, R_0$ and the locations $N_1, N_2$. These considerations apply to each pair of nucleation sites.

Two such simulated patterns for 3D arbitrarily located nucleation sites are shown in the top row of Fig.~\ref{fig:solidification}(a) and (b), and correspond to 3 and 4 nuclei, respectively. The final shape of the 3-sided and 4-sided voids are clear from the figure. These may be contrasted with corresponding experimental images showing analogous voids. Furthermore, the size of these voids can be predicted by suitably choosing the ratio $r/R_0$, according to the expected density difference between the liquid and solid phases. Voids with more sides were seldom observed in any of the powder particles.

These elementary considerations suggest that each facet on the surface observed in Fig.~\ref{fig:SEM_morphologies} is likely a separate grain and not a result of a cellular growth instability. The transition to dendritic growth could appear with each of these growing nuclei, and, given their near-equal growth rate, likely occurs simultaneously. This could explain the other two microstructures seen in Fig.~\ref{fig:SEM_morphologies}. 

\section{Discussion}
\label{sec:discussion}

Our results have clearly shown that the standard melting-solidification hypothesis---that particles are in the molten state as they exit the wheel-workpiece contact zone---requires the inclusion of an additional oxidation step to explain the experimental observations. Given the difficulty in accurately measuring chip temperatures, we obtained bounds using a detailed 3-body heat partition method as well as a microscopic machining model. Both these calculations strongly suggest that the chip does not reach anywhere near the melting temperature even under extreme cases. Consequently, melting must occur post chip ejection from the contact zone. Evaluation of oxidation kinetics and energetics showed that exothermic oxidation is the most likely cause for particle melting. This was also confirmed by cross-section analysis, which showed the occurrence of a thin oxide layer on the surface of the particles. Subsequently, the molten drop assumes a spherical shape and undergoes rapid solidification to form a single spherical particle. We now discuss some additional features of the process that couldn't be explored in the present manuscript. 

\subsection{Thermomechanical analysis of surface patterns}  
The three primary patterns that occur on the surface of the spherical particles (\emph{cf.} Fig.~\ref{fig:SEM_morphologies}) are typical of solidification structures observed during casting \cite{flemings1974solidification}. However, the precise thermo-mechanical conditions governing the development of each of these structures remains as yet unclear. The situation is complicated by multiple factors. Firstly, given the rapid timescales for solidification, the possibility of multiple simultaneous nucleation sites implies that the temperature field in the sphere is most likely not radially symmetric. This then necessitates the use of a fully numerical solution capable of solving multiple, coupled moving boundary Stefan problems in a confined geometry. Furthermore, the solution must also be able to handle rapid cooling rates $10^5$ $^\circ$C/s and non-equilibrium effects such as the formation of BCT metastable phases. As per our knowledge, such a comprehensive framework does not exist at present. The geometric model used in Sec.~\ref{subsec:solidification} circumvented this restriction by considering only the kinetics of multi-nucleus growth. This may be extended using a probabilistic description for the nucleation process \cite{Markworth1988} to provide additional analytical estimates. However, as we alluded to in Sec.~\ref{subsec:solidification}, it is quite likely that a more detailed evaluation of something as basic as the competition between multiple-nucleation sites and the spherical--cellular front transition requires is still in order. 

Finally, the timescale for solidification also mandates the use of non-equilibrium phase diagrams to account for the formation of meta-stable phases \cite{levi1982heat, galenko1997local}. This makes the problem significantly more complex and not particularly amenable to analytical treatment at present. Consequently, standard interface instability analyses \cite{mullins1964stability, ben1984pattern, alexandrov2021review} as well as paraboidal tip solutions \cite{ivantsov1988temperature} are no longer strictly valid and cannot, for instance, explain morphology transitions like the ones seen in Fig.~\ref{fig:SEM_morphologies}. We are presently evaluating a modified diffuse interface model and hope to present detailed investigations of these solidification patterns in a future manuscript. 

\subsection{Melting-resolidification \emph{versus} abrasion-burnishing for spherical particle formation}

At the outset, mention was made of two parallel mechanisms for the formation of spherical particles in abrasion, \emph{cf.} Fig.~\ref{fig:sphere_applications}. It is imperative that we now comment on when each of these mechanisms are operative. Firstly, it is clear that for the melting--resolidification mechanism, the need for either local melting (as with large-scale, extreme pressure contacts) or an additional reaction mechanism (as with oxidation) is necessary to bring material locally into the molten state. Furthermore, for the melting-resolidification route, perfectly spherical shapes of the intermediate liquid drops, such as the ones shown in Fig.~\ref{fig:SEM_images}, are governed by the so-called Bond number Bo defined as
\begin{equation}
  \text{Bo} = \left(\frac{R}{\lambda_c}\right)^2
\end{equation}
where $R$ is the drop radius and $\lambda_c = \sqrt{\gamma/h\rho}$ is the capillary length for the metal. Perfect spherical shape is possible only when Bo $\ll 1$ so that surface tension effects dominate over gravity (or any other convective effects for instance). For the particles observed and reported in this work, $R \sim 50\,\mu$m and $\lambda_c \sim 0.5$ cm so that Bo $\sim 10^{-2} \ll 1$. When such a condition is satisfied, perfect sphericity results and we can be certain that the melting-resolidification mechanism is operative.  

Secondly, in situations involving lubricated contacts (e.g., rolling fatigue), the possibility of either large local temperature or exposure to oxidation is remote. In these cases, it is almost certain that any spherical particles must form via the repeated mechanical action alone. While such action may, at first, appear far fetched, it is quite often observed in nature---the perfect rounding of pebbles in a river bed is one common instance. Here, the river flow imparts some directionality to the process so that ellipsoidal shapes result, yet in a completely isotropic situation, one can as well postulate the formation of nearly spherical final shapes. However, an explicit demonstration of this process, via suitable \emph{in situ} studies has hitherto not been undertaken, to the best of our knowledge.  

\subsection{Spherical abrasion particles as a source of powders}

In addition to its historical significance, the formation of spherical particles in the steel system is also of potential practical interest. Spherical powder particles form the primary stock material for metal additive manufacturing applications. Given their perfect spherical shape, one can then ask if these particles can be used as a starting point for powder-based consolidation processes, such as, for instance, powder metallurgy or metal additive manufacturing. We have conducted preliminary tests in this direction and our results appear to be very promising. If successful, it will perhaps link Robert Hooke and his monumental treatise with one of the most active areas of research in contemporary materials science, and perhaps not for the last time.

\section{Conclusions}
\label{sec:conclusions}

The formation of spherical powder particles in abrasion of steels poses an interesting scientific problem that we believe can be explained using the modified melting-solidification hypothesis presented in this manuscript. Heat partition models were used to obtain estimates for the exiting chip temperature and it was found that they were unlikely to reach anywhere near the melting point. Based on consideration of oxidation kinetics and EDS analysis of the surface of the final spherical particles, we established the need for an intermediate oxidation route. Energy balance arguments showed that even the oxidation of a thin surface layer of thickness $\sim 1\%$ of the sphere radius would release enough heat to melt the entire drop. TEM analysis of the final surface provided evidence for the occurrence of rapidly solidified BCT structures post-solidification. Finally, we postulated a simple kinematic description of surface morphology using purely geometric arguments. The complexity of the rapid solidification process include multiple nucleation sites, metastable phase formation and growth and morphological instability. We hope to address some of these problems in a forthcoming manuscript.     

\bibliographystyle{vancouver}
\bibliography{bibfile}
\end{document}